\documentstyle[12pt]{article}
\title{A Third-Quantized Approach to the Large-N Field Models}
\author{\large V.P.Maslov and O.Yu.Shvedov \\[3mm]
\em Sub-Department of Quantum Statistics and Field Theory,\\
\em Department of Physics, Moscow State University, \\
\em 119899, Vorobievy Gory, Moscow, Russia}
\date{July,16, 1998}
\begin{document}
\maketitle

\begin{flushright}
hep-th/9807134 \\
Talk presented at the 
International Conference \\
``Problems of Quantum Field Theory - 98'',\\
Dubna, Russia, July, 13-17, 1998
\end{flushright}

\begin{abstract}
Large-N field systems are considered from an unusual  point  of  view.
The Hamiltonian  is presented in a third-quantized form analogously to
the second-quantized  formulation  of  the  quantum  theory  of   many
particles. The   semiclassical   approximation   is   applied  to  the
third-quantized Hamiltonian.  The  advantages  of  this  approach   in
comparison with 1/N-expansion are discussed.
\end{abstract}
\footnotetext{e-mail: olshv@ms2.inr.ac.ru, shvedov@qs.phys.msu.su}

\newcounter{eqn}
\def\lab{\refstepcounter{eqn}\eqno(\arabic{eqn})}
\def\l#1{\lab\label{#1}}
\def\r#1{(\ref{#1})}

Different physical  systems  are  described  by different mathematical
equations. Classical mechanics is  formulated  in  terms  of  ordinary
differential equations  on  a finite time-dependent set of quantities.
Quantum mechanics,  classical field and string theories are associated
with partial  differential  equations on a function of a finite number
of arguments.  Qunatum field  theory  and  ``first-quantized''  string
theory are related with the Schrodinger equation on the functional.

Each subsequent  type  of  systems  (equations)  is in some sense more
complicated than the previous one.  For example, ordinary differential
equations are  much  more suitable for numerical analysis than partial
differential equations,  while functional Schrodinger equation is  too
complicated for  direct  numerical  analysis.  One can talk then about
classical, first-quantized and second-quantized systems.  There  is  a
correspondence between  these  types:  one  can  perform  quantization
procedure, as well as apply  semiclassical  approximation  (asymptotic
methods) for these systems (equations).

The notion   of   third  quantization  naturally  appears  in  quantum
cosmology \cite{1}
in investigation of  the  wormhole  transitions,  since  the
number of  universes can be variable.  The relationship between such a
multi-universe system and ordinary quantum field theory is the same as
the correspondence   between   quantum  and  classical  field  theory,
classical and quantum mechanics. An analogous situation also arises in
the field string theory.

Large-$N$ systems   are  much  more  complicated  than  the  small-$N$
systems. For  example,  system  of  a  large  number  of  differential
equations may  arise  as  a  lattice  approximation  for  the  partial
differential equation,  so that it is a very important result that  by
the WKB-technique  such  a  system  can  be  reduced  to  the ordinary
differential equation.

Further, quantum mechanics of a large number  of  particles  resembles
quantum field   theory,   since   one   can   introduce  creation  and
annihilation operators and  use  them  in  order  to  reformulate  the
quantum statistics  in  terms  of quantum field theory.  Semiclassical
approximation to this theory leads to  the  self-consistent  equations
such as  Hartree  or  Vlasov  equations,  so  that  large-$N$  quantum
mechanics can be reduced to the  nonlinear  quantum  dynamics  of  one
particle described by the Hartree equation.

This talk deals with the development of the analogous approach for the
large-$N$ quantum field theory. Analogously to the quantum statistics,
one can represent it via operators of creating and annihilating fields
and apply the semiclassical approximation. The corresponding classical
theory will resemble the quantum-field-theory  functional  Schrodinger
equation but will be nonlinear.

The notion of quasiparticles which can be created and annihilated even
in the system of a fixed number of particles playes an important  role
in quantum statistics,  superfluidity and superconductivity theory. It
is remarkable  that  semiclassical  equations for  the  third-quantized
large-$N$ theory  correspond  to  the  theory  with variable number of
fields. In this effective theory fields can be  created  (annihilated)
analogously to quasiparticles.

For the simplicity, consider the $O(N)$-symmetric scalar theory of $N$
fields. States of such a system at fixed moment of time are  specified
by functionals\\
$\Psi_N[\varphi_1(\cdot),...,\varphi_N(\cdot)]$,
the inner product is formally written by the functional integral\\
$
\int D\varphi_1 ... D\varphi_N
|\Psi_N[\varphi_1(\cdot),...,\varphi_N(\cdot)]|^2$,
while the Hamiltonian has the form,
$$
\begin{array}{c}
H_N = \int d{\bf x} \sum_{a=1}^N \left( -\frac{1}{2}
\frac{\delta^2}{\delta\varphi^a({\bf x})\delta\varphi^a({\bf x})}+
\frac{1}{2}\nabla\varphi_a({\bf x})\nabla\varphi_a({\bf x})+
\frac{m^2}{2}\varphi_a({\bf x})\varphi_a({\bf x})\right)+
\\
+\frac{\varepsilon}{4}
\sum_{a,b=1}^N \int d{\bf x} \varphi_a({\bf x})
\varphi_a({\bf x})\varphi_b({\bf x})\varphi_b({\bf x}).
\end{array}
\l{1}
$$
The third-quantized formulation is the following.  One should consider
the theory of random number of fields instead of the theory  of  fixed
number of fields. The states are specified by Fock vectors,
$
\left(
\begin{array}{c}
\Psi_0\\
\Psi_1[\varphi_1(\cdot)]\\
...\\
\Psi_N[\varphi_1(\cdot),...,\varphi_N(\cdot)]\\
...
\end{array}
\right),
$
while the Hamiltonian has a diagonal form,
$
\left(
\begin{array}{cccc}
H_0 & 0 & 0 & ... \\
0   & H_1 & 0 & ... \\
0 & 0 & H_2 & ...   \\
... & ... & ... &...
\end{array}
\right),
$
since the  number  of  fields  is  conserved.  If  one   considers   a
restriction on states to be symmetric under transpositions,
\\
$
\Psi_N[\varphi_1,...,\varphi_i,...,\varphi_j,...\varphi_N] =
\Psi_N[\varphi_1,...,\varphi_j,...,\varphi_i,...\varphi_N]
$,
one will be able   to   introduce   creation   (annihilation)   operators
$A^+[\varphi(\cdot)]$ ($A^-[\varphi(\cdot)]$)     transforming     the
$N$-field state to the $N+1$ ($N-1$) field state as follows,
$$
\begin{array}{c}
(A^+[\varphi(\cdot)]\Psi)_k[\varphi_1(\cdot),...,\varphi_k(\cdot)]=
\frac{1}{\sqrt{k}}\sum_{a=1}^k \delta(\varphi(\cdot)-\varphi_a(\cdot))
\times
\\
\times
\Psi_{k-1}[\varphi_1(\cdot),...,\varphi_{a-1}(\cdot),
\varphi_{a+1}(\cdot),...,\varphi_k(\cdot)],
\\
(A^-[\varphi(\cdot)]\Psi)_{k-1}
[\varphi_1(\cdot),...,\varphi_{k-1}(\cdot)]
= \sqrt{k} \Psi_k [\varphi(\cdot),
\varphi_1(\cdot),...,\varphi_{k-1}(\cdot)]
\end{array}
$$
analogously to  quantum  statistics.  The  Hamiltonian  \r{1}  can  be
presented as
$$
\begin{array}{c}
H= \int D\varphi A^+[\varphi(\cdot)] \int d{\bf x}
\left(
-\frac{1}{2}
\frac{\delta^2}{\delta \varphi({\bf x})
\delta \varphi({\bf x}) }+
\frac{1}{2}\nabla\varphi({\bf x})\nabla\varphi({\bf x})+
\frac{m^2}{2}\varphi({\bf x})\varphi({\bf x})\right)
A^-[\varphi(\cdot)]+
\\
+
\frac{\varepsilon}{4} \int D\varphi D\phi
\int d{\bf x} \varphi^2({\bf x}) \phi^2({\bf x})
A^+[\varphi(\cdot)]A^-[\varphi(\cdot)]
A^+[\phi(\cdot)]A^-[\phi(\cdot)].
\end{array}
$$
There are many ways to perform the semiclassical approxiamtion. One of
them is to  consider  the  Heisenberg  equations  of  motion  for  the
operators $A^{\pm}[t,\varphi(\cdot)] = e^{iHt} A^{\pm}[\varphi(\cdot)]
e^{-iHt}$,
$$
i\dot{A}^{\pm} = [A^{\pm},H]
\l{2}
$$
and substitute them by $c$-numbers.  However,  it is not clear if  one
can use  the  semiclassical conception since the canonical commutation
relations (CCR)
$$
[A^{\pm}[\varphi(\cdot)], A^{\pm}[\phi(\cdot)] ] =0,
[A^-[\varphi(\cdot)], A^+[\phi(\cdot)]]=\delta(\varphi(\cdot)-
\phi(\cdot))
\l{3}
$$
do not  contain  any  small  parameter.  However,  one  can  perform a
rescaling
$$
A^{\pm}\sqrt{\varepsilon} = \Phi^{\pm}
\l{4}
$$
which makes the CCR \r{3} containing $\varepsilon$. On the other hand,
Heisenberg equations for $\Phi^{\pm}$ do not depend on  $\varepsilon$,
since the  right-hand  side  of  eq.\r{2}  consists  of  a  linear  in
$A^{\pm}$ part being $\varepsilon$-independent and a cubic term  which
is proportional to $\varepsilon$. Therefore, by the substitution \r{4}
the small parameter $\varepsilon$ is moved from eq.\r{4} to CCR.

Although the   semiclassical   approximation    is    applicable    at
$\varepsilon\to 0$,  it  is not equivalent to the perturbation theory,
analogously to the soliton quantization method in QFT  which  is  more
general than the perturbation theory \cite{2}.

Substituting operators   $\Phi^{\pm}$   by (time-dependent)
$c$-numbers  $\Phi^*$  and
$\Phi$, one obtains the following classical equation,
$$
\begin{array}{c}
i\frac{d}{dt}\Phi_t[\varphi(\cdot)]=
\int d{\bf x}
\left(
-\frac{1}{2}
\frac{\delta^2}{\delta \varphi({\bf x})
\delta \varphi({\bf x}) }+
\frac{1}{2}\nabla\varphi({\bf x})\nabla\varphi({\bf x})+
\frac{m^2}{2}\varphi({\bf x})\varphi({\bf x})\right)
\Phi_t[\varphi(\cdot)]+
\\
+
\frac{1}{2}
\int d{\bf x} \varphi^2({\bf x})
\int D\phi
\phi^2({\bf x})
\Phi_t^*(\phi(\cdot))\Phi_t(\phi(\cdot))
\Phi_t(\varphi(\cdot))
\end{array}
\l{5}
$$
which is  a nonlinear functional Schrodinger equation corresponding to
the {\it nonlinear quantum}  field theory.

To construct   the   systematic  semiclassical  theory,  consider  the
following ansatz (cf.\cite{3}),
$$
\Psi^t_{\varepsilon}
=\exp\left(\frac{i}{\varepsilon}S^t\right)
U^{\varepsilon}_{\Phi^t} Y^t
\l{5*}
$$
corresponding to    the    transformation    shifting   the   creation
(annihilation) operators,
$$
U^{\varepsilon}_{\Phi}= \exp\left[
\frac{1}{\sqrt{\varepsilon}}\int D\phi
(\Phi(\varphi(\cdot))A^+(\varphi(\cdot))
-\Phi^*(\varphi(\cdot))A^-(\varphi(\cdot)))
\right]
$$
Substituting the     ansatz     to     the    Schrodinger    equation,
$i\dot{\Psi}^t_{\varepsilon}=H{\Psi}^t_{\varepsilon}$, one obtains the
expression for    the    real    number    $S^t$    in    the    order
$O({\varepsilon}^{-1})$; the classical equation \r{5} in the order
$O({\varepsilon}^{-1/2})$ and  the  following leading approximation for
the evolution equation,
$
i\dot{Y}^t = H_2 Y^t,
$
corresponding to the quadratic Hamiltonian $H_2$
$$
\begin{array}{c}
H_2= \int D\varphi A^+[\varphi(\cdot)] \int d{\bf x}
\left(-\frac{1}{2}\frac{\delta^2}{\delta \varphi({\bf x})
\delta \varphi({\bf x})}
+\frac{1}{2}(\nabla\varphi)^2({\bf x})
+ \frac{
m^2+(\Phi_t,\varphi^2({\bf x})\Phi_t)
}{2}\varphi^2({\bf x})
\right)
\times
\\
\times
A^-[\varphi(\cdot)]+
\frac{1}{4}
\int d{\bf x}
\left[
\int D\varphi \varphi^2({\bf x})
(
\Phi_t^*[\varphi(\cdot)]A^-[\varphi(\cdot)]
+ \Phi_t[\varphi(\cdot)]A^+[\varphi(\cdot)]
)
\right]^2
\end{array}
\l{6}
$$
containing the terms of the types $A^+A^+$,  $A^+A^-$,  $A^-A^-$.  One
can notice  that the effective time-dependent Hamiltonian \r{6} allows
the processes of creating and annihilating the fields,  analogously to
the quasiparticle Hamiltonian in statistics.

To construct  asymptotic  solutions  to  the  $N$-field equation,  one
should project expression \r{5*} on the  $N$-field  space.  Since  the
amplitude $P_N\Psi$  cannot  be  exponentially  small (otherwise,  the
constructed approximate solution will be much less than the accuracy),
it is necessary to mention that
$
N=\frac{1}{\varepsilon} \int D\varphi |\Phi[\varphi(\cdot)]|^2 + O(1)
$ (cf.\cite{3}).
One can notice that our approach is applicable if $\varepsilon\to 0$,
$N\to\infty$, $\varepsilon N \to const$ (these are the usual conditions
of applicability of the $1/N$-expansion).

Different semiclassical methods can be  applied  to  our  system.  For
example, one  can  quantize  \cite{4}
the  periodic  solutions of the classical
equation \r{5}    of    the     type
$\Phi[t,\varphi(\cdot)] = \Phi[\varphi(\cdot)] e^{-i\Omega t}$.
Quantization about periodic solutions is equivalent to the problem  of
diagonalization of the quadratic Hamiltonian \r{6}, which is reducable
to investigation of the variation system for eq.\r{5}.  This system is
obtained as   follows.  One  should  consider  the  system  containing
eq.\r{5} and    conjugated    equation.    Then    one     substitutes
$\Phi^*+\delta\Phi^*$ instead   of   $\Phi^*$   and  $\Phi+\delta\Phi$
instead of   $\Phi$.   The   small   variations   $\delta\Phi^*$   and
$\delta\Phi$ are  independent,  contrary  to $\Phi^*$ and $\Phi$.  The
corresponding linearized system for  $\delta\Phi^*$  and  $\delta\Phi$
which is  called  as  the  variation  system described the spectrum of
quantum fluctuations.  More precisely,
the solution of the  variation  system
with the     following     time     dependence,    $\delta\Phi    \sim
e^{i(\beta-\Omega)t}$, $\delta \Phi^* \sim e^{i(\beta+\Omega)t}$,
is associated with the excitation with energy $\beta$.

To remove the UV-divergences,  one should consider the regularization,
substitute the local field $\varphi({\bf x})$ by  the  cutoffed  field
$\varphi_{\Lambda}({\bf x})  =  \int d{\bf y} A_{\Lambda}({\bf x}-{\bf
y})\varphi({\bf y})$
 and consider the limit $A_{\Lambda}({\bf z})\to  \delta({\bf  z})$
($\lambda, m$  are  singular) in such a way that renormalized mass and
coupling constant should be finite.

Let us  compare  the  developed  approach  with  known  approaches  to
the large-$N$-expansion. \\
1. A usual approach is  resummation  of  Feynman  bubble  graphs,  or,
equivalently, calculations  of  the  functional  integral by using the
auxiliary field
\cite{5}.  However,  this approach allows us to  consider  only
small perturbations  around  vacuum  and  breaks down if the number of
particles $n$ is large, $n\sim N$ \cite{6}. \\
2. There is   also   a   collective-field   approach
\cite{7} based   on   the
$O(N)$-symmetric substitution  to  the   Schrodinger   equation.   Our
approach is more general,  since we consider not only $O(N)$-symmetric
states.\\
3. Averaging  the  second-quantized  Heisenberg  equations  on  fields
$\hat{\varphi}_a$, one can obtain the equation  which  is  similar  to
eq.\r{5} \cite{8}.
However, the semiclassical theory based on eq.\r{6} canot be
obtained by using this procedure.

The developed approach can be generalized to other  models.  First  of
all, one can consider the system of $N$ fields $\varphi_a$ interacting
with the scalar field $\chi$.  One can perform the third  quantization
of the  field  $\varphi$ and treat $\chi$ as a second-quantized field.
The classical phase space consists then of the sets
$\{ (\chi({\bf  x}),\pi({\bf x}),\Psi[\varphi(\cdot)]\}$ of
the field   $\chi$,   momentum   $\pi$    and    complex    functional
$\Psi[\varphi(\cdot)]$.
Using this observation,  one can  investigate  the  nonsymmetric  with
respect to  transpositions  wave  functionals:  one  considers  fields
$\varphi_1,...,\varphi_k$ as     second-quantized      fields      and
$\varphi_{k+1},...,\varphi_N$ as        third-quantized        fields.
Generalizations to the fermi-case is also trivial.

We conclude the talk by the following observations.

1. When one constructs the quantum field theory,  one always considers
the Hamiltonian  to  be  a linear operator.  However,  we see that the
nonlinear quantum field theory \r{5} seems to  be  in  agreement  with
relativistic invariance  and  other  principles of QFT,  because it is
considered as a corollary of the linear theory of $N$ fields which  is
beleived to possess all necessary properties.

2. One  always considers the quantum field theories of fixed number of
fields: the set of elementary particles is fixed and does  not  depend
on the   state.   However,  the  Hamiltonian  \r{6}
corresponding to variable number of fields
should  obey  all
principles of QFT since it is also a corollary af the large-$N$ theory.

\end{document}